\documentclass[twocolumn, journal]{IEEEtran}

\def\BibTeX{{\rm B\kern-.05em{\sc i\kern-.025em b}\kern-.08em
    T\kern-.1667em\lower.7ex\hbox{E}\kern-.125emX}}
\usepackage{amssymb,amsmath,color,graphicx,subfigure}
\usepackage[noadjust]{cite}
\usepackage{algorithm}
\usepackage{algorithmic}
\usepackage{bm}
\usepackage{stfloats}
\usepackage{subfigure}
\usepackage{booktabs}
\usepackage{tabularx}

\begin{document}

\title{Spectrum Sharing for 6G Integrated Satellite-Terrestrial Communication Networks Based on NOMA and Cognitive Radio}

\author{Xin Liu, Kwok-Yan Lam, Feng Li, Jun Zhao, Li Wang
\thanks
{
%This work was supported by the National Natural Science Foundations of China  under Grant
%U1833102 and the Natural Science Foundation of Liaoning Province under Grants
%2019-ZD-0014 and 2020-HYLH-13.
(\emph{Correspondence author: Feng Li})
}
\thanks{X. Liu is with the School of Information and Communication Engineering, Dalian University of Technology, Dalian 116024, China (e-mail:liuxinstar1984@dlut.edu.cn).}
\thanks{K. Lam and J. Zhao are with School of Computer Science and Engineering, Nanyang Technological University, 639798, Singapore. (kwokyan.lam@ntu.edu.sg, junzhao@ntu.edu.sg)}
\thanks{F. Li is with School of Information and Electronic Engineering, Zhejiang Gongshang University, Hangzhou, 310018, China. F. Li is also at School of Computer Science and Engineering, Nanyang Technological University, 639798, Singapore. (fengli2002@yeah.net)}
\thanks{L. Wang is with College of Marine Electrical Engineering, Dalian Maritime University, Dalian, 116026, China. (liwang2002@dlmu.edu.cn)}
}

\maketitle

\begin{abstract}
The explosive growth of bandwidth hungry Internet applications has led to the rapid development of new generation mobile network technologies that are expected to provide broadband access to the Internet in a pervasive manner. For example, 6G networks are capable of providing high-speed network access by exploiting higher frequency spectrum; high-throughout satellite communication services are also adopted to achieve pervasive coverage in remote and isolated areas. In order to enable seamless access, Integrated Satellite-Terrestrial Communication Networks (ISTCN) has emerged as an important research area. ISTCN aims to provide high speed and pervasive network services by integrating broadband terrestrial mobile networks with satellite communication networks. As terrestrial mobile networks began to use higher frequency spectrum (between 3GHz to 40GHz) which overlaps with that of satellite communication (4GHz to 8GHz for C band and 26GHz to 40GHz for Ka band), there are opportunities and challenges. On one hand, satellite terminals can potentially access terrestrial networks in an integrated manner; on the other hand, there will be more congestion and interference in this spectrum, hence more efficient spectrum management techniques are required. In this paper, we propose a new technique to improve spectrum sharing performance by introducing Non-orthogonal Frequency Division Multiplexing (NOMA) and Cognitive Radio (CR) in the spectrum sharing of ISTCN. In essence, NOMA technology improves spectrum efficiency by allowing different users to transmit on the same carrier and distinguishing users by user power levels while CR technology improves spectrum efficiency through dynamic spectrum sharing. Furthermore, some open researches and challenges in ISTCN will be discussed.
\end{abstract}

%\begin{IEEEkeywords}
%air-ground integrated networks; multi-UAV deployment; UAV base station; UAV relay
%\end{IEEEkeywords}

\IEEEpeerreviewmaketitle
\section*{\textcolor[rgb]{0.00,0.18,0.51}{\textbf{Introductions}}}

In order to enable pervasive network connectivity, Integrated Satellite-Terrestrial Communication Networks (ISTCN) has emerged as an important research area. ISTCN aims to provide high-speed and pervasive network services by integrating broadband terrestrial mobile networks with satellite communication networks. The ISTCN can provide reliable communications and global interconnections for disaster affected areas, remote areas and emergency areas, where the terrestrial communication facilities are not easy to use \cite{JiaM}.

The future 6G networks are expected to offer unprecedented opportunities for Smart Cities and Internet of Things applications through their global seamless coverage, G-bit communication capacity, ultra reliable real-time communications and ubiquitous machine type communications. This new generation terrestrial mobile network achieved their functionalities by intelligently and optimally exploiting the higher frequency spectrum, typically in the range of 3GHz to 40GHz.
However, for suburban and isolated geographic locations, the coverage of high-speed terrestrial mobile networks could be limited hence need to be complemented by satellite communications in order to meet the connectivity requirements of safety critical applications such as Internet of Vehicles and Industry 4.0 control systems \cite{Chattopadhyay}.

As terrestrial mobile network began to use higher frequency spectrum which overlaps with that of satellite communications (e.g. 4GHz to 8GHz for C band and 26GHz to 40GHz  for Ka band), there are vast opportunities as well as difficult challenges. On one hand, satellite terminals can potentially access terrestrial network in an integrated manner; on the other hand, there will be more congestion and interference in this spectrum, hence more efficient spectrum management techniques are required.
The objective is to make full use of the complementary advantages of satellite networks and terrestrial mobile networks, so as to realize the all-weather and all-regional seamless coverage of high-speed mobile broadband network .

In addition, it aims to effectively alleviate the shortage of satellite spectrum resources by applying spectrum sharing technology to reuse the terrestrial spectrum for the satellite communications \cite{FengB}. In 6G mobile communications, Non-orthogonal Multiple Access (NOMA) and Cognitive Radio (CR) are two most promising spectrum sharing technologies \cite{Islam}.
NOMA is different from the traditional Orthogonal Multiple Access (OMA), which uses non-orthogonal resource allocation approach to accommodate more users \cite{Zhang}. At the transmitter, the transmit information of multiple users is superimposed and encoded in the power domain by intentionally adding the interference information. At the receiver, Successive Interference Cancellation (SIC) is used to separate the user information by sequentially detecting and canceling the signal of each user. It is estimated that NOMA can improve the current spectrum efficiency by 5$\sim$15 times \cite{Mounchili,ZhouF}. Therefore, the satellite-terrestrial NOMA spectrum sharing can make one satellite frequency band accommodate more users and thus greatly improve the communication capacity. CR, based on software radio, allows the system to adaptively adjust transmit parameters by sensing the current communication environment, so as to achieve efficient  spectrum resource utilization \cite{LiuXin,LiF}. It can share the spectrum resources among the heterogeneous communication systems through spectrum sensing and dynamic reconfiguration capability. As a secondary user (SU), the CR system can opportunistically utilize the idle spectrum of primary user (PU) or share the spectrum with the PU at a lower power \cite{Hattab,ZouJ}. Satellite-terrestrial CR spectrum sharing makes the satellite system and terrestrial system utilize the same spectrum resources, which can alleviate satellite spectrum tension effectively.

However, if compared with the spectrum sharing studies for terrestrial networks, the related works for ISTCN still remain insufficient. In \cite{YanX}, the capacity of NOMA-uplink satellite network was analyzed, which has proved the advantage of NOMA to improve the satellite communication capacity. In \cite{JiaoJ}, a joint resource allocation and network stability optimization was proposed to maximize the long-term network utility of NOMA-downlink satellite system. In \cite{Sagduyu}, regret minimization solution was put forward for PU and SU's spectrum access to the satellite resources when existing cognitive interferers. In \cite{Chae}, cooperative transmission strategy was proposed for cognitive satellite networks, where the mobile users in the terrestrial network can help the communication of the satellite network to improve its transmission performance. Nevertheless, NOMA and CR for integrated satellite-terrestrial spectrum sharing are less considered.

In this article, NOMA and CR based spectrum sharing for the ISTCN is proposed to solve the problem of satellite spectrum scarcity. The contributions of the article are concluded as follows. (1) The network model and network access model for the ISTCN are proposed, which allow the satellite system and terrestrial system share the same spectrum by the integration of satellite and terrestrial components; (2) The satellite-terrestrial NOMA spectrum sharing is presented to let multiple users to access the same satellite spectrum by superposition coding in power domain; (3) The satellite-terrestrial CR spectrum sharing is proposed to make the satellite system and terrestrial system share the same spectrum resource by suppressing their mutual interferences; (4) By combining NOMA and CR, the satellite-terrestrial CR-NOMA spectrum sharing is put forward to achieve full spectrum access by using both the idle and busy spectrum.

\section*{\textcolor[rgb]{0.00,0.18,0.51}{\textbf{Integrated Satellite-Terrestrial Communication Networks}}}

\begin{figure*}[ht]
\centering
\includegraphics[width=5in]{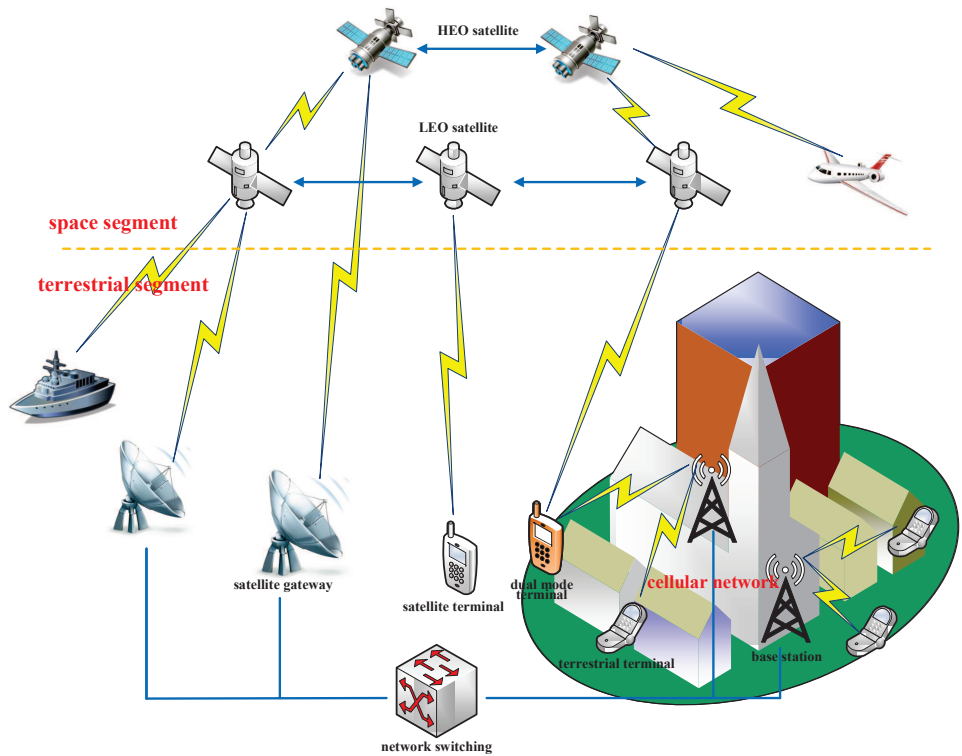}
\centering
\caption{Network model for ISTCN.}
 \label{fig:ISTCN}
\end{figure*}

Contemporary satellite communication services are no longer competitors of terrestrial cellular network. Instead, they are often adopted to complement cellular network services so as to provide seamless coverage. Terrestrial cellular networks are suitable for areas with high user density in the urban areas; however, they are typically less cost effective in covering remote and even isolated geographic areas. While satellite communications can provide large area coverage at low cost, they have their limitations in covering urban areas due to the influence of shadowing effect. Therefore, ISTCN is believed to be a suitable approach to achieve the global coverage with optimal cost. The network model for the ISTCN is shown in Fig. \ref{fig:ISTCN}, which is an integrated satellite-terrestrial system composed of one or more Highly-Elliptical-Orbit (HEO) and Low-Earth-Orbit (LEO) satellites and terrestrial cellular system. Both the terrestrial system and satellite system operate in the same frequency band to ensure the global seamless coverage of the user terminals. In the ISTCN, the satellite terminal and terrestrial terminal can communicate with each other depending on the network switching between the satellite system and cellular system, and also a dual-mode satellite terminal can choose either of the two systems to communicate by measuring the transmission cost.

%%%%%%%%%%%%%%%%%%%%%%%%%%%%

\begin{figure*}[ht]
\centering
\includegraphics[width=5in]{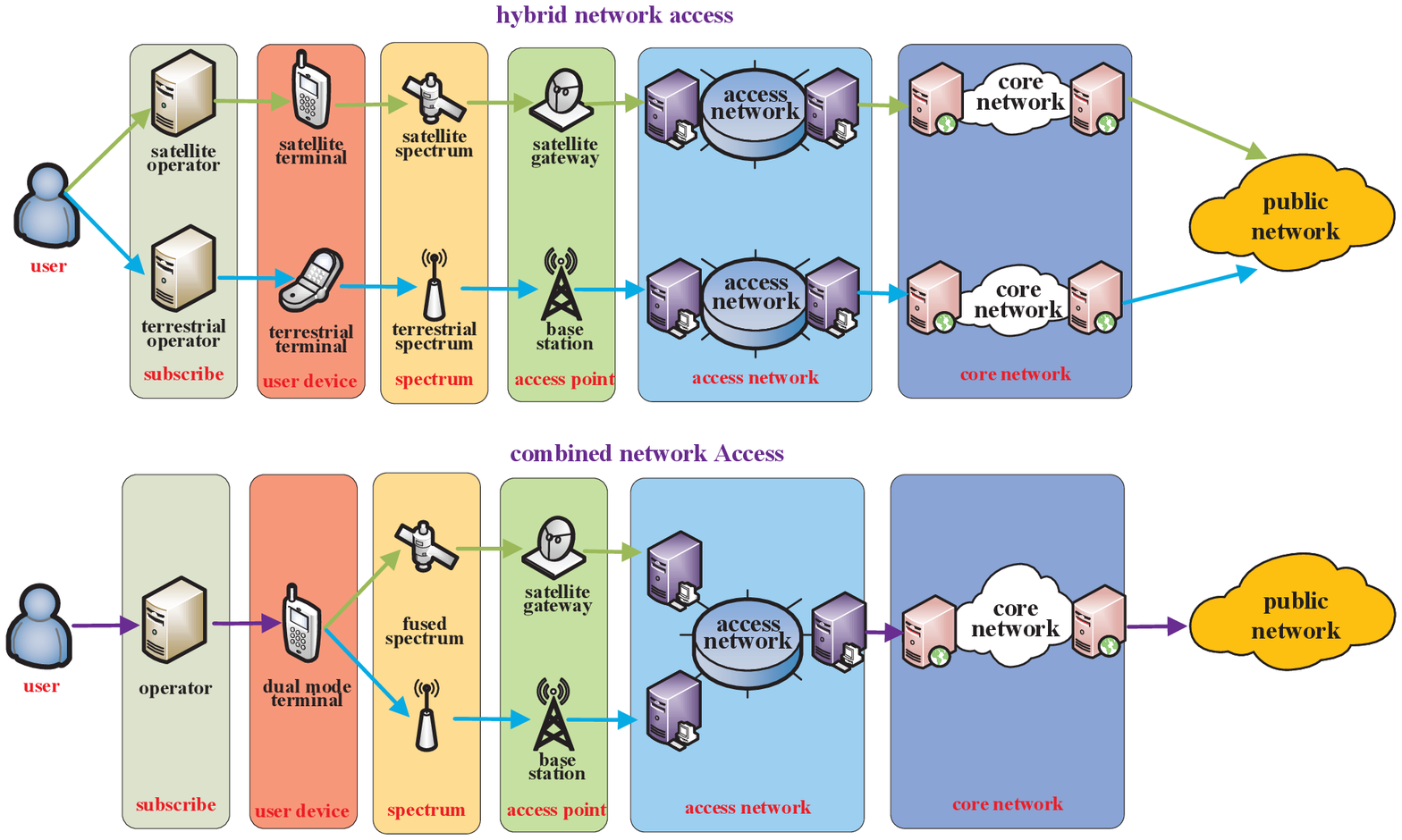}
\centering
\caption{Network access mode for ISTCN.}
 \label{fig:Access_mode}
\end{figure*}

The network access model for spectrum sharing in ISTCN includes hybrid network access (HNA) and combined network access (CNA), as shown in Fig. \ref{fig:Access_mode}. In the HNA, the user terminal and subscription are different, and the access network and core network of each system are disjoint and linked together through the public network. Therefore, the users may have good access to the two systems, but there is no integration between them. The satellite system and terrestrial system can adopt the same or different air interface technology depending on the specific network scenario.
In the HNA, however, the user only have one terminal and one subscription, and the services of the two systems are almost seamless switching. Therefore, the quality of service (QoS) of HNA is higher due to the system integration. But the two systems have to adopt compatible air interface technology and share the same frequency band.
%%%%%%%%%%%%%%%%%%%%%%%%%%

In the ISTCN, the satellite system and terrestrial system may coexist in the same frequency band to alleviate the satellite spectrum scarcity. The existing spectrum sharing methods are mainly divided into the following two categories.

\textbf{Static spectrum sharing}: The idea of static spectrum sharing is spatial isolation, which can reuse the time, frequency and other resources through orthogonal access manner in different spatial areas. However, it allocates the specific spectrum resource for each user, which can not meet the dynamic spectrum demands of the users, resulting in that the load of some spectrum is too heavy while other spectrum has higher idle rate. In 6G, NOMA, as a new static spectrum sharing approach based on power domain multiplexing, has been  proposed to allocate the same time-frequency resource to different users, which can greatly improve the spectrum efficiency compared with 5G.

\textbf{Dynamic spectrum sharing}: By adopting CR technology, the communication system can opportunistically use the underutilized frequency resources to achieve better dynamic spectrum management. Satellite-terrestrial CR spectrum sharing can realize the heterogeneous integration of satellite network and terrestrial network and solve the problem of satellite spectrum shortage.

Therefore, NOMA and CR as efficient spectrum sharing technologies can make the ISTCN achieve the interconnections between massive satellite terminals and terrestrial  terminals under the limited satellite spectrum resources.
\section*{\textcolor[rgb]{0.00,0.18,0.51}{\textbf{Satellite-Terrestrial NOMA Spectrum Sharing}}}
The core idea of NOMA is to realize the multi-user multiplexing of single time-frequency resource block by introducing a new power domain dimension. At the transmitter, the signals of different users are set with different power levels, which are transmitted in the same resource block by superposition coding. While at the receiver, the signals of different users are separated and decoded by using SIC in the descending order of the power levels. .
\subsection*{\textcolor[rgb]{0.00,0.18,0.51}{\textbf{NOMA Spectrum Sharing Model}}}
\begin{figure*}[ht]
\centering
\includegraphics[width=5in]{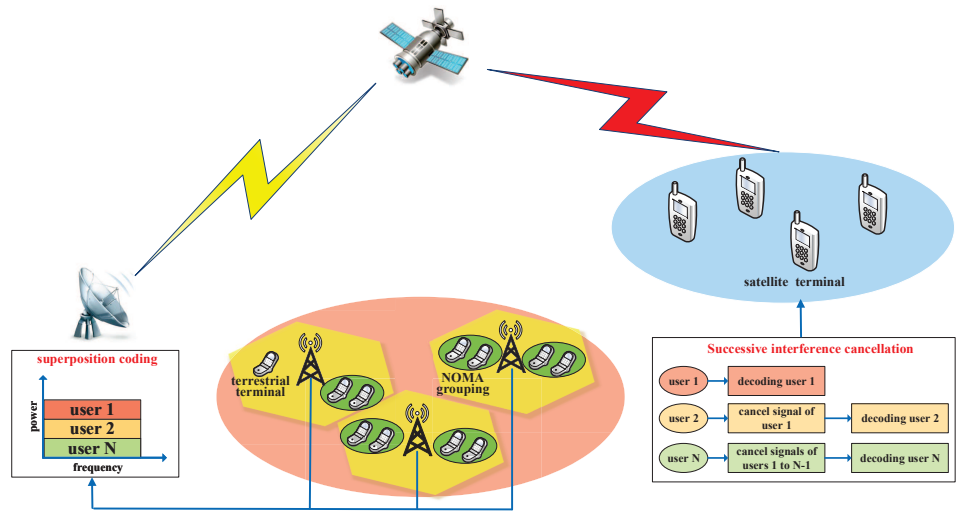}
\centering
\caption{NOMA based satellite-terrestrial spectrum sharing model.}
 \label{fig:Sa_NOMA}
\end{figure*}

The satellite-terrestrial NOMA spectrum sharing model is shown in Fig. \ref{fig:Sa_NOMA}, where the terrestrial users access the satellite spectrum through NOMA.
In the satellite uplink, the data and channel state information (CSI) of the terrestrial terminals are sent to the satellite gateway, which then groups the users according to the CSI and allocates the maximum transmit power to each user for superposition coding. The signals of the same grouping users are sent to the satellite in the same frequency band. The satellite receiver uses SIC to decode the signals of each user. If a user transmits stronger signal in a better link, its signal will be decoded first. And the signal in a poor link will be decoded from the remaining signals after subtracting the decoded signals by SIC.
In the satellite downlink, the satellite transmitter allocates the power of each user according to the link quality, and the user in a poor link is allocated larger power to ensure the receiving performance. The NOMA signal is transmitted to each satellite terminal, which uses SIC to first decode the signals with larger power and then decode its own signal from the remaining signals.
%%%%%%%%%%%%%%%%%%%%%%%%%%%%
\subsection*{\textcolor[rgb]{0.00,0.18,0.51}{\textbf{User NOMA Grouping}}}
The terrestrial users can be divided into several groups, each of which is assigned a separate frequency band for NOMA transmission. The NOMA grouping can reduce the multi-user interference in decoding by decreasing the number of users in the same frequency band. The advantage of NOMA is obvious only when the users with great channel differences are assigned to one group. In the terrestrial NOMA, the physical distance between user and base station is used as the grouping basis, whereby the center user and edge user within the coverage of base station are usually assigned to one group. However, the satellite communication channel is more complex than the terrestrial mobile communication channel, whose path attenuation is not sensitive to the user's geographical location. Therefore, the distance grouping basis is no longer applicable in the satellite communications. Satellite NOMA grouping needs to fully consider other attenuation characteristics of the satellite channels besides the free space loss, such as beam gain, shadow fading, multipath fading, and rain fading etc. The channel fading difference of different satellite users can be used as the grouping basis to eliminate the insensitive path attenuation.
\subsection*{\textcolor[rgb]{0.00,0.18,0.51}{\textbf{Cooperative Satellite-Terrestrial NOMA}}}
Cooperative NOMA is mostly used in the downlink of a communication system, whereby the user with good channel can help to decode the information of the user with poor channel, so as to enhance its receiving performance. In the ISTCN, cooperative NOMA can be carried out among different satellite and terrestrial terminals to improve the transmission performance of the users in fading satellite channels. As shown in Fig. \ref{fig:Co_NOMA}, the satellite terminals  in  fading channels can form a NOMA group with either the satellite terminals in good channels or the terrestrial terminals. The satellite transmits the NOMA signal to the satellite terminals and the terrestrial base station. The satellite terminals in good channels first decode all the signals by SIC, and then use decode-and-forward (DF) protocol to send the decoded signals to the satellite terminals in fading channels. However, the terrestrial terminals cannot achieve the prior information of the satellite terminals and thus are unable to decode their signals. Therefore, the terrestrial terminals first decode their own signals and subtract the decoded signals from the received signal, and then use amplify-and-forward (AF) protocol to send the remaining signal to the satellite terminals in fading channels.
\begin{figure*}[ht]
\centering
\includegraphics[width=5in]{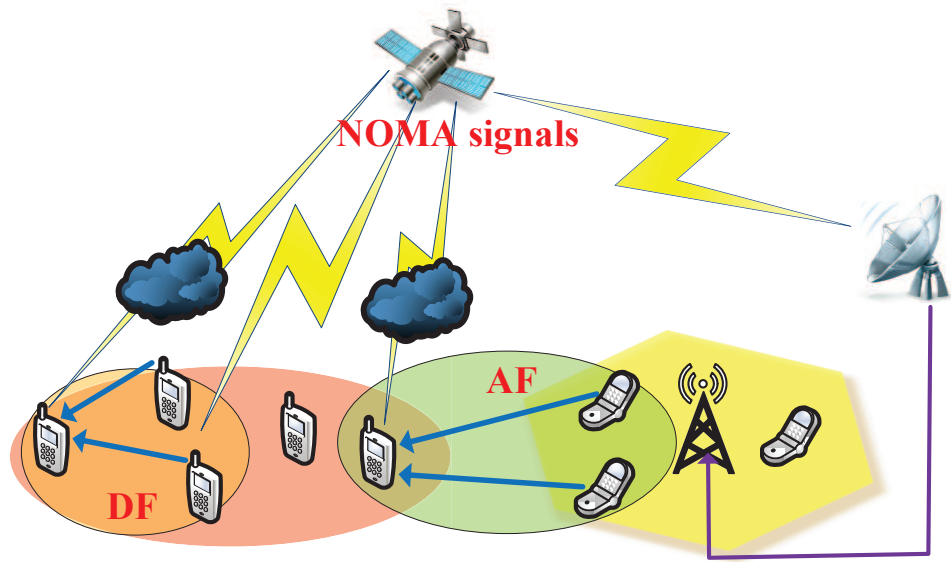}
\centering
\caption{NOMA-based cooperative spectrum sharing.}
 \label{fig:Co_NOMA}
\end{figure*}
%%%%%%%%%%%%%%%%%%%%%%%%
\section*{\textcolor[rgb]{0.00,0.18,0.51}{\textbf{Satellite-Terrestrial CR Spectrum Sharing}}}
\subsection*{\textcolor[rgb]{0.00,0.18,0.51}{\textbf{CR Spectrum Sharing Model}}}
Using CR technology, the satellite communication system can flexibly share spectrum resources with the terrestrial  communication system, which can improve the spectrum utilization via opportunistically accessing the frequency bands permitted by the licensed users.  The typical CR spectrum sharing scenarios of the ISTCN can be divided into two categories. One is licensed satellite system and terrestrial CR system, and the other is licensed terrestrial system and satellite CR system.

\begin{figure*}[ht]
\centering
\includegraphics[width=5in]{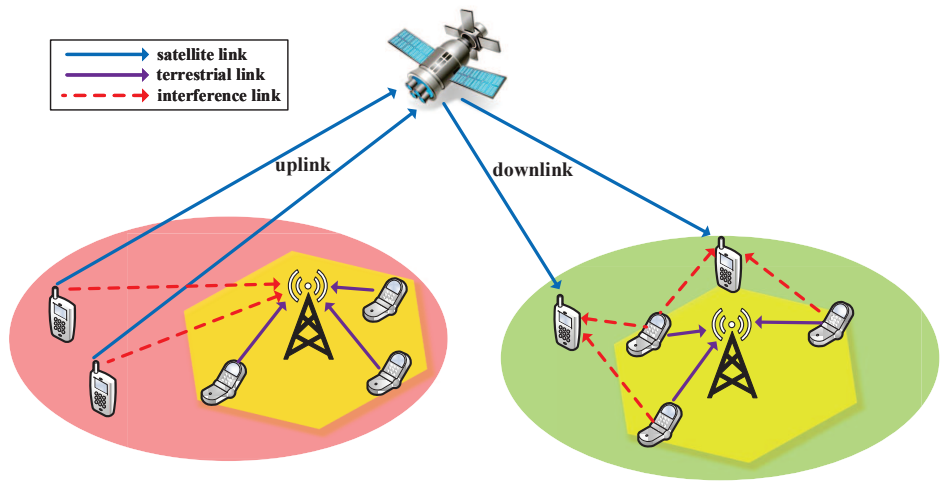}
\centering
\caption{CR spectrum sharing model.}
 \label{fig:Sa_CR}
\end{figure*}

As shown in Fig. \ref{fig:Sa_CR}, in the satellite uplink, the satellite CR system communicates in the terrestrial channels by spectrum sharing technology. If the terrestrial user is not using the channel, the satellite user can transmit data with its maximum power. However, the satellite user must always sense the channel state. If the presence of the terrestrial user in the channel has been detected, the satellite user has to switch to another idle channel. However, if there is no idle channels, the satellite user may continue to use this channel but cause harmful interference to the terrestrial system. In the satellite downlink, the interference from the terrestrial user will also decrease the satellite communication performance. To achieve low-interference spectrum sharing, the satellite user must detect the spectrum occupation state of the terrestrial system accurately and select an idle channel for transmissions. In addition, the satellite system can also access the busy spectrum by controlling its transmit power so that its power does not exceed the maximum interference tolerated by the terrestrial system.
\subsection*{\textcolor[rgb]{0.00,0.18,0.51}{\textbf{Interference suppression technology}}}
The premise of CR spectrum sharing is that the interference between the satellite system and terrestrial system does not affect their normal communications. Some interference suppression technologies are introduced as follows.

\textbf{Interference cognition}: Interference cognition can detect the interference holes in the surrounding electromagnetic environment, identify the interference and estimate the channel quality, which can provide the basis for the anti-interference decision. In the terrestrial communication, the interference mostly occurs in the channel from CR system to licensed system, and the interference cognition is usually defined around the licensed receiver. However, there may be two-way interference in the ISTCN, and the interference cognition should be defined both around CR system and licensed system. For example, the interference cognition for the ISTCN can be defined around the earth station and the satellite spot beam.

\textbf{Power control}: CR system combines channel state, receiver signal-to-noise ratio (SNR) and interference information to flexibly adjust its transmit power  to avoid interference with licensed user in the same frequency band. On the one hand, the transmit power can be minimized to save the energy of the satellite terminal on the premise of ensuring the communication capacity. On the other hand, the power can be optimized to maximize the communication capacity of the ISTCN providing that the  interference threshold is not exceeded.

\textbf{Satellite beamforming}: Beamforming technology  can transmit the target signal along the desired direction by weighting the received signals of  antenna array elements. Beamforming allows multiple users to utilize the same frequency band in the same geographical area at the same time, which makes it possible to deploy dense networks with less interference in the unexpected direction. It can be used as an interference cancellation technology in the transmitter or receiver of the satellite, which can realize the spectrum sharing of satellite system and terrestrial system in the angle domain.
\section*{\textcolor[rgb]{0.00,0.18,0.51}{\textbf{Satellite-Terrestrial CR-NOMA Spectrum Sharing}}}
CR can realize the spectrum coexistence of satellite system and terrestrial system, while NOMA can achieve the sharing access of limited satellite spectrum by massive users. Therefore, by combining CR and NOMA, the satellite spectrum utilization can be further improved.
Satellite terminals can access both the idle and busy spectrum via CR-NOMA, which will achieve high-efficient full spectrum access.

\textbf{CR-NOMA in idle spectrum}: Multiple satellite terminals can access the idle spectrum by NOMA, which will not bring any interference to the terrestrial system.
However, the available idle frequency bands are usually discontinuous and fragmented, which are difficult to meet the broadband access of massive satellite users. Spectrum aggregation technology has been put forward to aggregate discrete idle frequency bands into broadband spectrum to support large-bandwidth data transmissions. Non-continuous Orthogonal Frequency Division Multiplexing (NC-OFDM) can realize the subcarrier aggregation by zeroing the non-idle subcarriers according to their bandwidth and locations. Therefore, by introducing NC-OFDM into NOMA, the
superimposed broadband signal can be transmitted over the aggregated idle subcarriers.

\textbf{CR-NOMA in busy spectrum}: Satellite terminal and terrestrial terminal can share the same spectrum by NOMA, but interfere with each other. To guarantee the terrestrial communication performance, the satellite signals are first decoded with the terrestrial signals as the noise. Then the decoded satellite signals are cancelled from the received NOMA signal by SIC, and the remaining signal is used to decode the terrestrial signals without the interference from the satellite. However, if the transmit power of terrestrial terminals is large enough or the terrestrial communication performance is ignored, the terrestrial signals can be first decoded to guarantee the decoding performance of satellite signals.
%%%%%%%%%%%%%%%%%%%%%%%%%%%%
\begin{figure*}[htbp]
\centering
\includegraphics[width=6in]{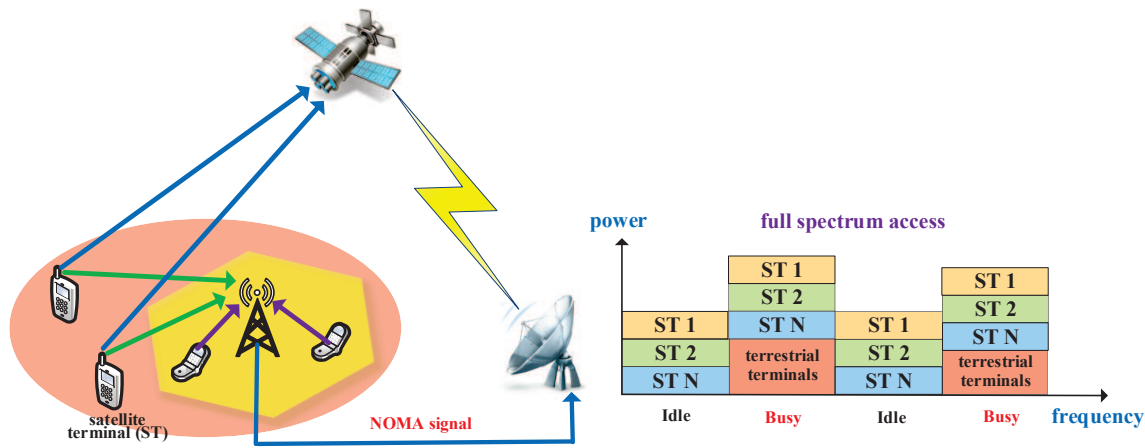}
\centering
\caption{CR-NOMA spectrum sharing.}
 \label{fig:Co_NOMA}
\end{figure*}

Though the CR-NOMA can make the ISTCN achieve full spectrum access, the multi-user interference caused by NOMA and the satellite-terrestrial interference cause by CR may decrease the NOMA decoding performance. The satellite terminals and terrestrial terminals should be grouped appropriately and the decoding order must be properly arranged according to the power and service requirements of the users.
\section*{\textcolor[rgb]{0.00,0.18,0.51}{\textbf{Open Researches and Challenges}}}
This article has introduced some fundamental works on spectrum sharing for ISTCN, such as ISTCN network model, satellite-terrestrial NOMA spectrum sharing, satellite-terrestrial CR spectrum sharing and satellite-terrestrial CR-NOMA spectrum sharing. However, there are still some open researches and challenges to be discussed in the future.

\textbf{Satellite spectrum sensing}: The premise of satellite-terrestrial spectrum sharing without interference is to perform accurate spectrum sensing. Due to the  limited satellite transmission capacity and the significant signal attenuation caused by atmospheric effects such as shadowing and rain fading, it is a great challenge to accurately sense satellite spectrum. In addition, the spectrum sensing in LEO satellite communication also faces the problems of mobility and available frequency shortage.

\textbf{Fair satellite NOMA grouping}: Beam edge users are located in the overlapping area of different satellite beams and will suffer great inter-beam interference. Therefore, multi-user interference and inter-beam interference will seriously reduce the decoding performance of the edge users. It is necessary to propose a fair satellite NOMA grouping method to allocate low-power users and fewer users for the groups of edge users, so as to decrease the NOMA decoding interference.

\textbf{Satellite NOMA receiver design}: Due to the shadowing, multipath fading, rain fading and other channel interference of satellite communication, it is difficult to carry out perfect SIC at the satellite terminal. When restructuring and canceling the inaccurate decoded signals, the decoding error will be transferred to the subsequent signal demodulation, which will decrease the decoding performance. Therefore, to guarantee SIC performance, the satellite receiver design should adopt some new signal processing technology to suppress the interference and noise, such as adaptive filtering, wavelet transform and weak signal detection etc.

\textbf{Integrated satellite-6G network}: 6G can support high-capacity, multi-service and high-speed wireless communications. Integrated satellite-6G network can meet the global coverage of mobile Internet and the ubiquitous network access of all kinds of users. To better integrate with the terrestrial 6G network, the satellite segment needs to reuse all the functional modules of 6G core network, such as Internet interface, quality of service, user mobility and security etc.

\section*{\textcolor[rgb]{0.00,0.18,0.51}{\textbf{Conclusion Remarks and Future Works}}}
In this article, we propose NOMA and CR based spectrum sharing schemes for ISTCN to improve the satellite spectrum utilization by allowing the satellite communication to share the spectrum licensed to the terrestrial communication. The satellite-terrestrial spectrum sharing need to sense the terrestrial spectrum state and suppress the mutual interference between satellite system and terrestrial system. Some interference suppression technologies for satellite-terrestrial spectrum sharing are also introduced. By combining CR and NOMA, the ISTCN can use CR-NOMA to achieve full spectrum sharing by accessing both the idle and busy spectrum. Finally, some promising researches and challenges for the ISTCN have been discussed.


\begin{thebibliography}{10}

\bibitem{JiaM}
M.~{Jia}, X.~{Gu}, Q.~{Guo}, W.~{Xiang}, and N.~{Zhang}, ``Broadband hybrid
  satellite-terrestrial communication systems based on cognitive radio toward
  5g,'' {\em IEEE Wireless Communications}, vol.~23, no.~6, pp.~96--106, 2016.

\bibitem{FengB}
B.~{Feng}, H.~{Zhou}, H.~{Zhang}, G.~{Li}, H.~{Li}, S.~{Yu}, and H.~{Chao},
  ``Hetnet: A flexible architecture for heterogeneous satellite-terrestrial
  networks,'' {\em IEEE Network}, vol.~31, no.~6, pp.~86--92, 2017.

\bibitem{Islam}
S.~M.~R. {Islam}, M.~{Zeng}, O.~A. {Dobre}, and K.~{Kwak}, ``Resource
  allocation for downlink noma systems: Key techniques and open issues,'' {\em
  IEEE Wireless Communications}, vol.~25, no.~2, pp.~40--47, 2018.

\bibitem{Chattopadhyay}
A.~{Chattopadhyay}, K.~{Lam}, and Y.~{Tavva}, ``Autonomous vehicle: Security by
  design,'' {\em IEEE Transactions on Intelligent Transportation Systems},
  vol.~to appear, pp.~1--15, 2020.

\bibitem{Zhang}
H.~{Zhang}, Y.~{Qiu}, K.~{Long}, G.~K. {Karagiannidis}, X.~{Wang}, and
  A.~{Nallanathan}, ``Resource allocation in noma-based fog radio access
  networks,'' {\em IEEE Wireless Communications}, vol.~25, no.~3, pp.~110--115,
  2018.

\bibitem{Mounchili}
S.~{Mounchili} and S.~{Hamouda}, ``Pairing distance resolution and power
  control for massive connectivity improvement in noma systems,'' {\em IEEE
  Transactions on Vehicular Technology}, vol.~69, no.~4, pp.~4093--4103, 2020.

\bibitem{ZhouF}
F.~{Zhou}, Y.~{Wu}, R.~Q. {Hu}, Y.~{Wang}, and K.~K. {Wong}, ``Energy-efficient
  noma enabled heterogeneous cloud radio access networks,'' {\em IEEE Network},
  vol.~32, no.~2, pp.~152--160, 2018.

\bibitem{LiuXin}
X.~{Liu}, S.~{Hu}, M.~{Li}, and B.~{Lai}, ``Energy-efficient resource
  allocation for cognitive industrial internet of things with wireless energy
  harvesting,'' {\em IEEE Transactions on Industrial Informatics}, pp.~1--1,
  2020.

\bibitem{LiF}
F.~{Li}, K.~{Lam}, X.~{Li}, Z.~{Sheng}, J.~{Hua}, and L.~{Wang}, ``Advances and
  emerging challenges in cognitive internet-of-things,'' {\em IEEE Transactions
  on Industrial Informatics}, vol.~16, no.~8, pp.~5489--5496, 2020.

\bibitem{Hattab}
G.~{Hattab} and M.~{Ibnkahla}, ``Multiband spectrum access: Great promises for
  future cognitive radio networks,'' {\em Proceedings of the IEEE}, vol.~102,
  no.~3, pp.~282--306, 2014.

\bibitem{ZouJ}
J.~{Zou}, H.~{Xiong}, D.~{Wang}, and C.~W. {Chen}, ``Optimal power allocation
  for hybrid overlay/underlay spectrum sharing in multiband cognitive radio
  networks,'' {\em IEEE Transactions on Vehicular Technology}, vol.~62, no.~4,
  pp.~1827--1837, 2013.

\bibitem{YanX}
X.~{Yan}, H.~{Xiao}, K.~{An}, G.~{Zheng}, and S.~{Chatzinotas}, ``Ergodic
  capacity of noma-based uplink satellite networks with randomly deployed
  users,'' {\em IEEE Systems Journal}, vol.~14, no.~3, pp.~3343--3350, 2020.

\bibitem{JiaoJ}
J.~{Jiao}, Y.~{Sun}, S.~{Wu}, Y.~{Wang}, and Q.~{Zhang}, ``Network utility
  maximization resource allocation for noma in satellite-based internet of
  things,'' {\em IEEE Internet of Things Journal}, vol.~7, no.~4,
  pp.~3230--3242, 2020.

\bibitem{Sagduyu}
Y.~E. {Sagduyu}, Y.~{Shi}, A.~B. {MacKenzie}, and Y.~T. {Hou}, ``Regret
  minimization for primary/secondary access to satellite resources with
  cognitive interference,'' {\em IEEE Transactions on Wireless Communications},
  vol.~17, no.~5, pp.~3512--3523, 2018.

\bibitem{Chae}
S.~H. {Chae}, C.~{Jeong}, and K.~{Lee}, ``Cooperative communication for
  cognitive satellite networks,'' {\em IEEE Transactions on Communications},
  vol.~66, no.~11, pp.~5140--5154, 2018.

\end{thebibliography}
\end{document}